\documentclass[12pt]{iopart}

\usepackage{iopams} 
\newcommand{\dd}{\mathrm{d}}
\usepackage{subcaption}
\usepackage{graphicx}
\usepackage{xcolor}
\begin{document}
\title[Theory and Instrumentation of Fourier Transform Spectroscopy]{Theory and Instrumentation of Fourier Transform Spectroscopy}
\author{Sohrab Sheikh-Sofla and Mohammad Neshat
}

\address{School of Electrical and Computer Engineering, 
College of Engineering, 
University of Tehran, 
Tehran, Iran}

\eads{\mailto{sheikh.sofla2016@ut.ac.ir}, \mailto{mneshat@ut.ac.ir}}

\begin{abstract}
Fourier transform spectroscopy (FTS) has been widely used as an analytical tool for many applications in science and engineering. In this paper, we describe the operation principle and practical implementation of an FTS prototype. First, the structure of an FTS setup based on optical interferometers is introduced, and its optical and electronic components are identified. A simple theory is presented to show the Fourier transform relationship between the measured interferogarm signal at the detector and the spectral intensity of the source. Then, simulation results based on ray-tracing method are demonstrated to show the effect of coherent length of the source on the spectral intensity. Finally, various practical considereations to implement an FTS setup including noise/interference reduction are discussed, and the effect of alignment on the measurement accuracy is investigated.
Since the design and implementation of an FTS measurement setup requires the knowledge from various parts of physics and engineering education,   therefore, we believe that FTS can provide  a  very comprehensive  educational tool for the last  year undergraduate  and graduate  students.
\end{abstract}

\section{\label{introduction}Introduction}

Spectroscopy is a measurement technique to study the interaction of the electromagnetic spectrum with matter. In spectroscopy, an electromagnetic (EM) wave interacts with a sample material under study, and the spectrum of the transmitted or reflected wave is measured by a spectrometer. The spectrum of the interacted wave with the sample material is normalized to the spectrum of the no sample measurement in order to obtain the absorption spectrum of the sample material. From absorption spectrum, one can obtain information about the structure and constituents of the sample material. For example, many biological materials, such as DNA, have absorption signatures in the terahertz band that can be studied by spectroscopy measurements \cite{globus2006terahertz}. 

There are various spectroscopy methods, among which the Fourier transform spectroscopy (FTS) has been widely in use for many applications. FTS is able to detect minute quantities in the ppm and ppb ranges, and to characterize materials, such as silicon wafers, in order to extract their thickness, oxygen content, and phonon modes \cite{mollmann2013fourier}. FTS is a versatile analytical method for characterizing both liquid and solid chemicals. In solid-state samples, attenuated total reflection of powdered samples such as microalgae biomass and polysaccharides has been demonstrated \cite{heise2013recent}. For liquid mixtures in the concentration range between 3 to 104 ppm, the transmittance spectra of some materials like soaked Nafion has been studied in \cite{bunkin2018dynamics,blue2010creating}. Accuracy of optical alignments can increase using fiber optics in FTS structure as shown in \cite{kellerer2017coherence}. Terahertz radiation (also called T-rays) can be employed for spectroscopy and imaging, from the laboratory to industry such as biological applications \cite{zouaghi2013broadband}. THz time-domain spectroscopy (TDS) is the most widely used technique. THz-TDS is advantageous at lower frequencies under 3 THz, while Fourier transform spectroscopy works better at frequencies above 5 THz \cite{han2001direct}.  

The structure of an FTS measurement setup is based on an interferometer such as Fabry-Perot, Lamellar grating, and Michelson configurations \cite{iwata2006f,ferhanoglu2009lamellar,griffiths2007fourier}. In an FTS, the spectrum of the interacting wave is obtained from the the Fourier transform of an interferogram signal. The interferogarm signal is the output of a power detector that is used to detect the power of the interference between the interacting wave with its delayed copy over different optical path lengths \cite{bell2012introductory}. In a Fourier transform spectroscopy setup, a wideband source such as a thermal lamp can be used to generate EM waves within a range of wavelengths in order to measure the absorption spectrum simultaneously for all wavelength at once. One of the main advantages of the simultaneous measurement at different wavelengths, like in FTS, is the so called Fellgett's advantage which results in the improvement of the signal-to-noise ratio as compared to an equivalent scanning monochromator  \cite{bell2012introductory}. 

Given a wide range of studies over various applications of FTS reported in the literature, and the lack of sufficient information on the instrumentation of an FTS setup, this paper focuses on important practical aspects so as to design, simulate, and implement a prototype of an FTS setup suitable for educational purposes. Since the design and implementation of an FTS measurement setup involves using the knowledge and skills from different topics taught in physics such as wave propagation, interference, optical detection, optical alignment techniques, signal analysis, control theory for precise mechanical movements and more, therefore, it provides a very comprehensive educational tool to solidify the knowledge and skills of last year undergraduate or graduate students in the aforementioned topics. In this paper, we describe a successful project in which one of the co-authors has used such knowledge and skills during the course of his master studies to design, simulate and implement an FTS system prototype with inexpensive laboratory items.

\section{Interferometers}
As mentioned before, an interferometer is the main component of an FTS setup. Here, we compare and contrast the most widely used interferometers for FTS application.

 \begin{figure}
  \begin{subfigure}[b]{0.55\textwidth}

   \includegraphics[width=1\textwidth,trim=0px 0px 0px 50px,clip=true]{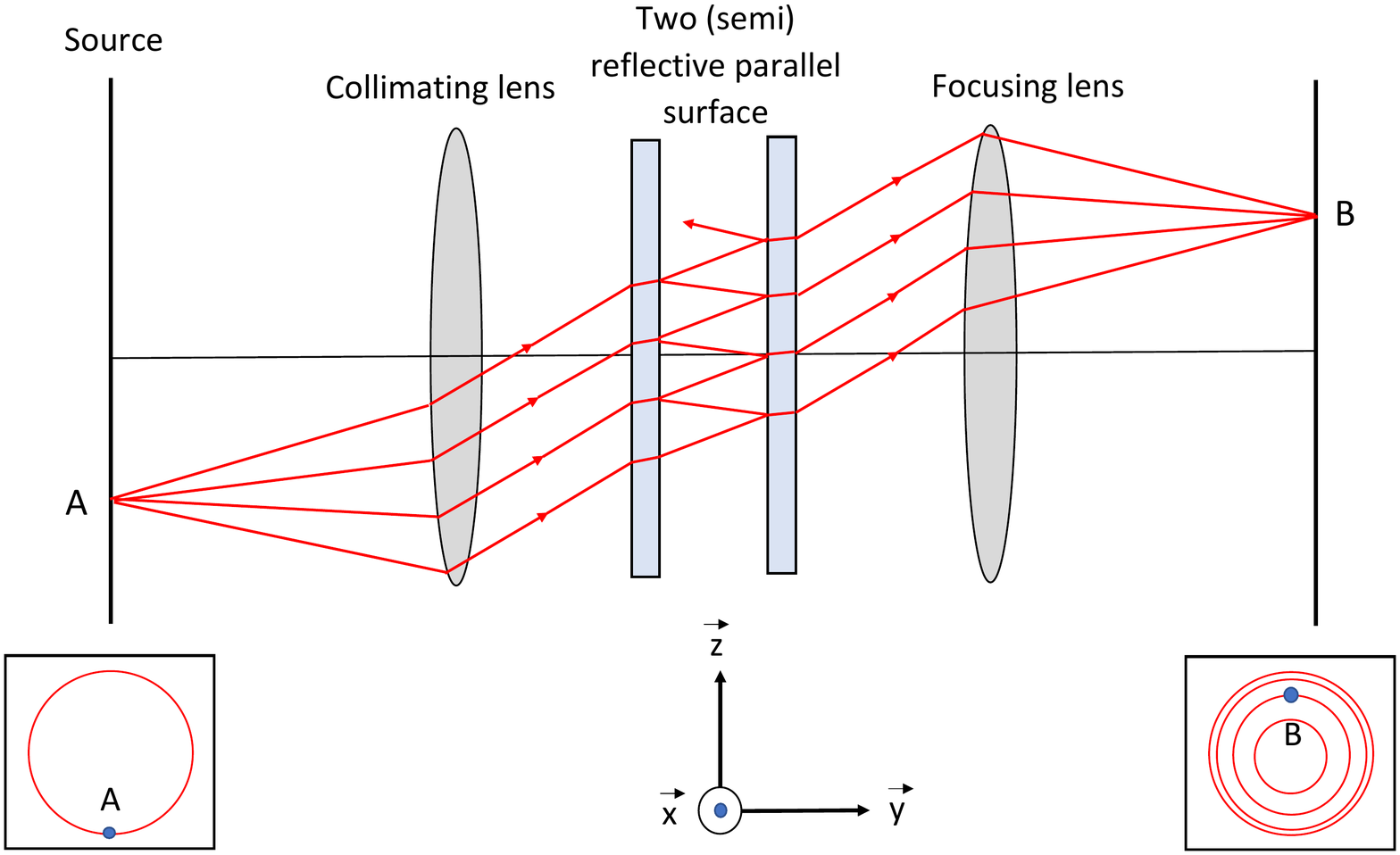}
    \caption{\label{fig1.3} }

    \end{subfigure}~
    \begin{subfigure}[b]{0.45\textwidth}

  \includegraphics[width=1\textwidth,trim=0px 120px 0px 120px,clip=true]{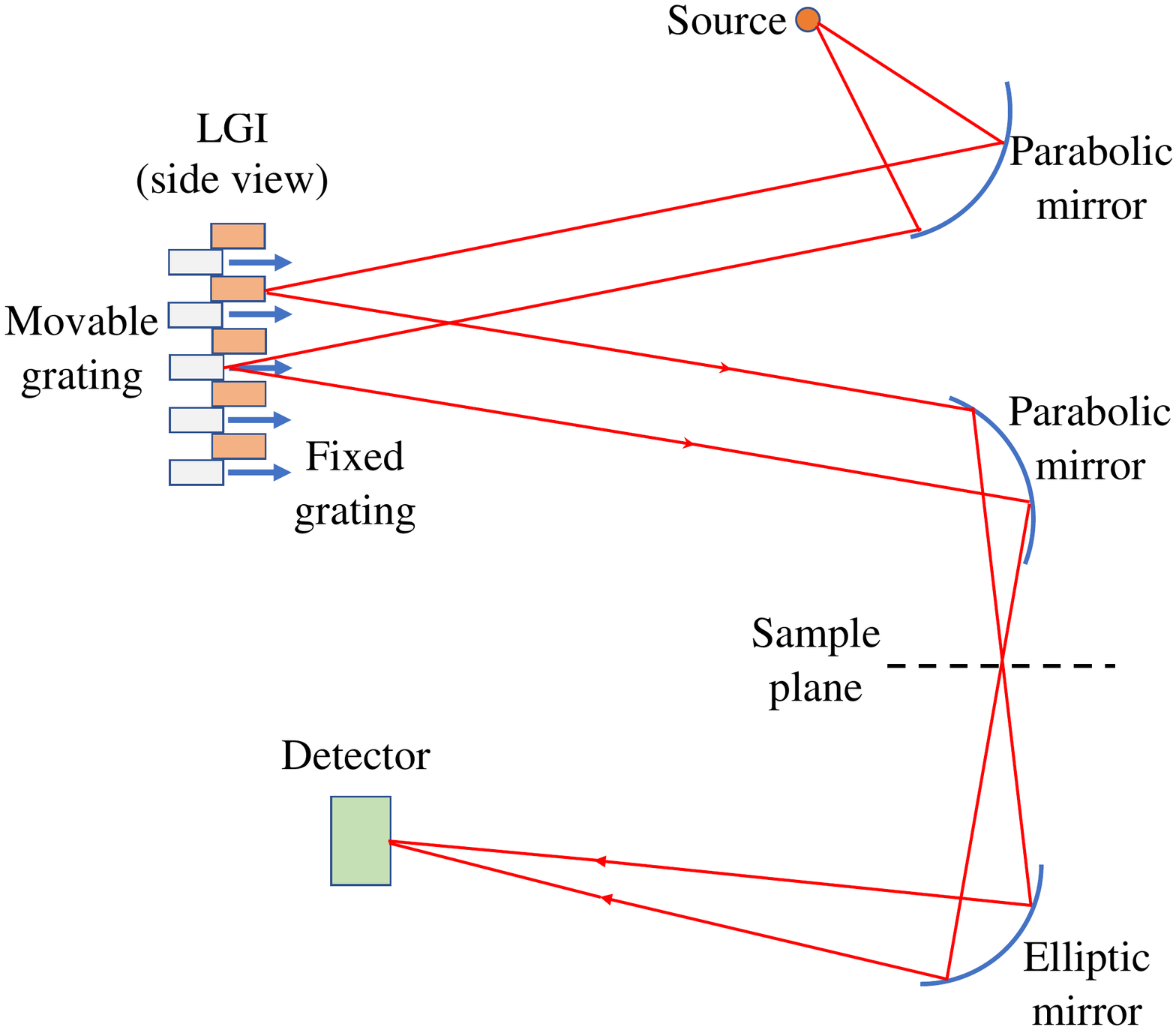}
    \caption{\label{fig1.2}}
    
    \end{subfigure}
   
    \begin{center} 
   \begin{subfigure}[b]{0.7\textwidth}

   \includegraphics[width=1\textwidth,trim=100px 0 0 70px,clip=true]{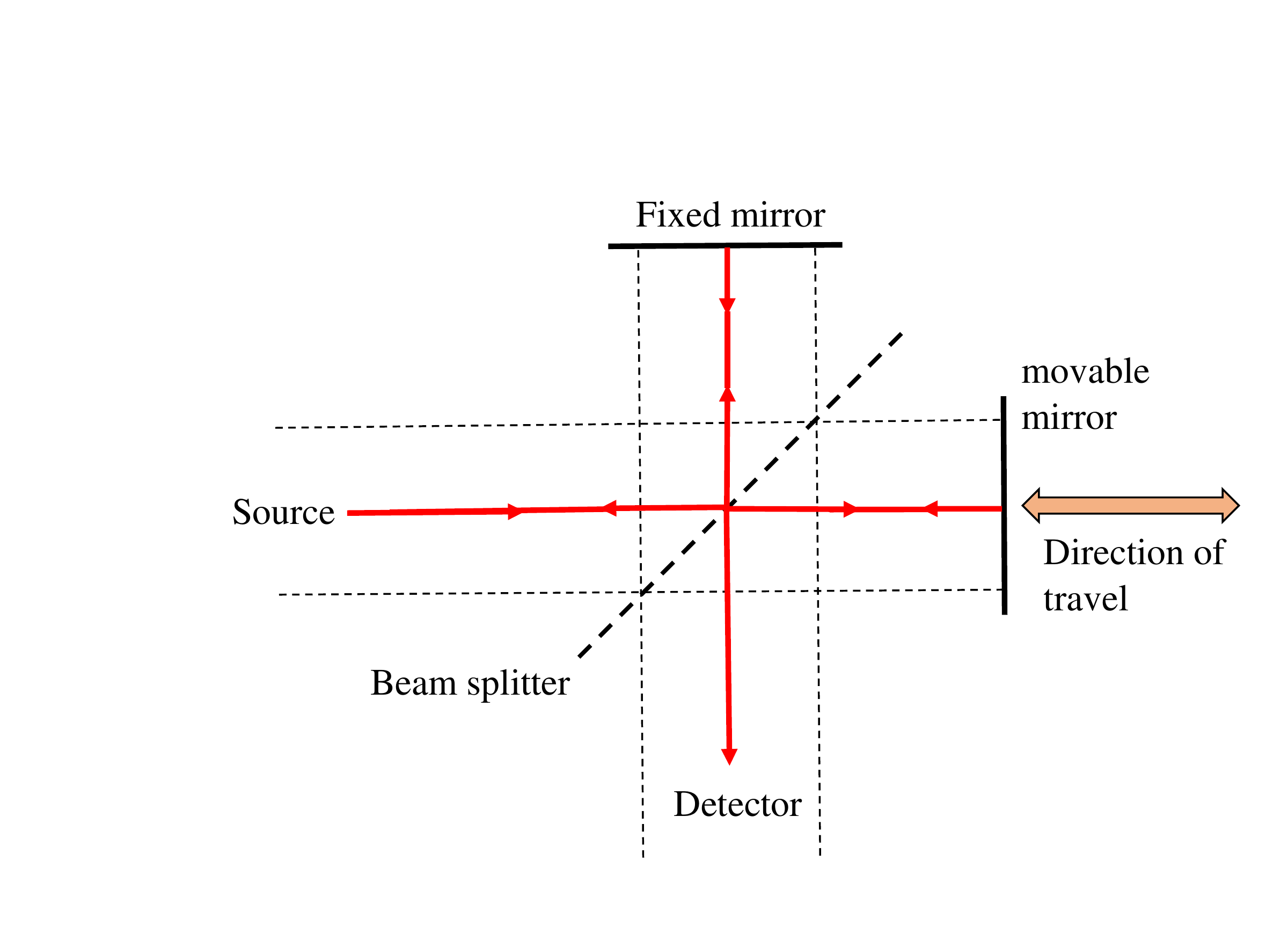}
   \caption{\label{fig1.1} }

    \end{subfigure}
 \end{center}
 \caption{\label{fig1}(a) Fabry-Perot etalon and equal-inclination interference fringes, (b) Lamellar grating structure, (c) Michelson intereferometer structure.}
 \end{figure}
 
\subsection{\label{Fabry-Perot}Fabry-Perot interferometer}
Fabry-Perot interferometer uses the phenomenon of multiple-beam reflection from the boundaries of a cavity bounded by two (semi)reflective parallel surfaces as shown in figure \ref{fig1}(a). Each time the wave encounters one of the surfaces, a portion of it is transmitted out of the cavity, and the remaining part is reflected back. The transmitted waves interfere with each other at the detector.
To make an interferogram with this structure, one needs to rotate the cavity around the x-axis to have interfered waves with different path lengths. One advantage of this interferometer is its simple implementation and easy alignment as compared to other interferometers. However, there are some drawbacks, e.g. the limitted angle of rotation of the cavity can restrict the spectral resolution, and there is a constraint on the achievable spectral bandwidth \cite{hernandez1988fabry}.

\subsection{\label{Lamellar}Lamellar grating interferometer}
In the Lamellar grating interferometer, waves interfere by reflecting from a grating structure with variable depth as shown in figure \ref{fig1}(b). In this structure, even though there are sufficient spectral resolution, but it needs a powerful source \cite{loewen2018diffraction}. Broadband powerful sources in some frequency bands such as in THz range are costly and not easily accessible.

\subsection{\label{Michelson}Michelson interferometer }
Conventional Fourier transform spectrometers work based on Michelson interferometer as shown in figure \ref{fig1}(c). It consists of one beam splitter (BS) and two mirrors. One of the mirrors can move back and forth to create different optical path lengths. The input wave emitting from the source is divided into two beams by the BS, where they propagate toward fixed and moving mirrors in two perpendicular arms. Upon reflection from the mirrors, the two beams are combined by the BS, and their interference power are measured by a detector. The interferogram is obtained by recording the output signal of the detector while one of the mirrors is moving. In the next Section, it will be shown that the spectral power density of the input wave is calculated by taking the Fourier transform of the interferogram. One advantage of the Michelson interferometer is that it loses less power than does other interferometers, and so it performs better in the frequency bands for which there are not low cost powerful sources. Moreover, it provides high spectral resolution, low aberration, fast scan, and broadband spectrum for each scan. However, it has some drawbacks such as difficult alignment, and high sensitivity to noise and vibration \cite{griffiths2007fourier}.

\section{Theory of FTS}
In the Michelson interferometer, the divided beams take the pass $z_1$ and $z_2$ in each arm, and  interfere at the detector position. The total electric field at the detector can be calculated as

\begin{eqnarray}
\label{eq.1}
    \mathbf{E}_R(z_1,z_2,\sigma)=rt\mathbf{E}_0(\sigma)[e^{-i(2\pi\sigma z_1)}+e^{-i(2\pi\sigma z_2)}] 
    \end{eqnarray}
\noindent where $\sigma$ is the wavenumber, i.e. the reciprocal of the wavelength $\sigma=\frac{1}{\lambda}$, $\mathbf{E}_0(\sigma)$ is the electric field amplitude of the source, $z_1$ and $z_2$ are distances that the two beams travel round-trip between the beam splitter and  mirrors,  $r$ and $t$ are the reflection and transmission coefficients of the beam splitter, respectively. The intensity of the total field within a differential spectral interval d$\sigma$ is given by

    \begin{eqnarray}
\label{eq.2}
\mathbf{I}(z_1,z_2,\sigma)\dd\sigma&=\mathbf{E}_R(z_1,z_2,\sigma) \mathbf{E}_R^*(z_1,z_2,\sigma)\dd \sigma\\
    \nonumber
    &=2\mathbf{E}_0^2(\sigma)|rt|^2(1+\cos{[2\pi(z_1-z_2)\sigma])}\dd \sigma
\end{eqnarray}

\noindent where the superscript $*$ denotes the complex conjugate. Denoting the optical path difference between the two beams in Michelson interferometer by $\delta=(z_1-z_2)$, the total intensity over all spectrum is calculated as 

\begin{eqnarray}
\label{eq.3}
\mathbf{I}_R(\delta)&=\int_{0}^{\infty}\!\mathbf{I}(\delta,\sigma)\dd\sigma\\
    \nonumber
    &=2|tr|^2\int_{0}^{\infty}\!\mathbf{E}_0^2(\sigma)\dd\sigma+2|tr|^2\int_{0}^{\infty}\!\mathbf{E}_0^2(\sigma)\cos{[2\pi\delta\sigma]}\dd\sigma
\end{eqnarray}

\noindent where $\mathbf{I}_R(\delta)$ is the interferogram that can be easily measured, and $\mathbf{E}_0^2(\sigma)$ can be defined as the input spectral intensity, $\mathbf{I}_0(\sigma)$. Rearranging equation (\ref{eq.3}) yealds

\begin{eqnarray}
\label{eq.4}
[\mathbf{I}_R(\delta)-\frac{1}{2}\mathbf{I}_R(0)]=2|tr|^2\int_{0}^{\infty}\!\mathbf{I}_0(\sigma)\cos{[2\pi\delta\sigma]}\dd\sigma
\end{eqnarray}

As it is evident in equation (\ref{eq.4}), the interferogram is related to the Fourier cosine transform of the input spectral intensity, therefore, one can simply obtain $\mathbf{I}_0(\sigma)$ by inverse Fourier cosine transform of the interferogram as

\begin{eqnarray}
\label{eq.5}
\mathbf{I}_0(\sigma)=\frac{1}{|rt|^2\pi}\int_{0}^{\infty}\![\mathbf{I}_R(\delta)-\frac{1}{2}\mathbf{I}_R(0)]\cos{[2\pi\delta\sigma]}\dd\delta
\end{eqnarray}

\section{Simulation results}
\begin{figure}
\begin{center}
  \includegraphics[width=0.5\textwidth]{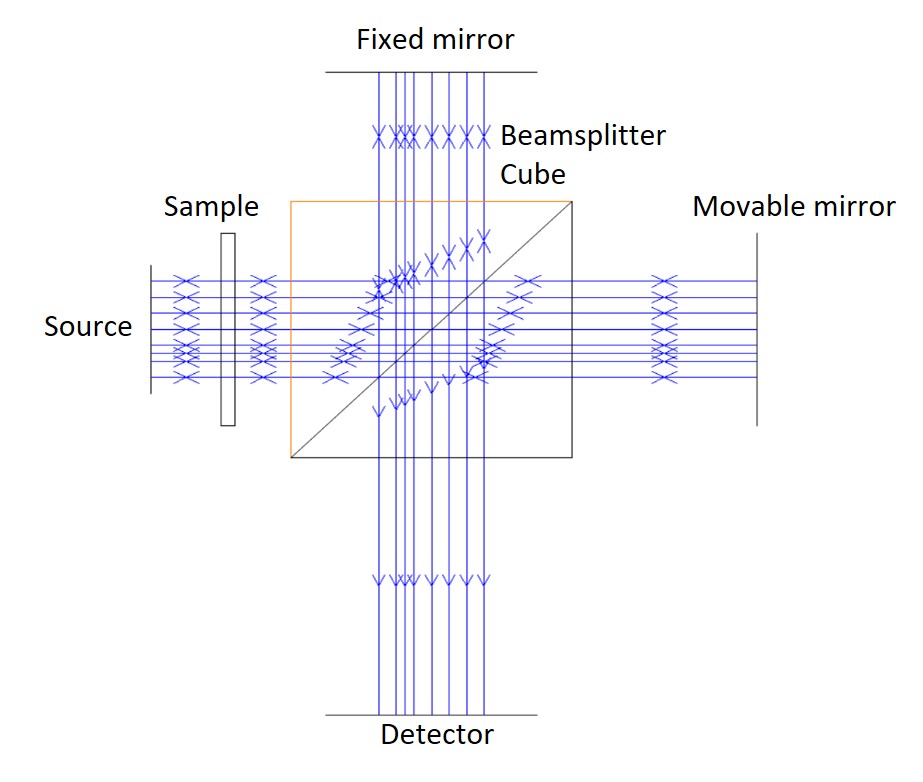}
\caption{\label{fig2}Simulated Michelson intereferometer structure in Zemax OpticStudio.}
\end{center}
\end{figure}

In this section, we present the simulation results of an FTS setup. Due to the large geometry of the setup as compared to the operation wavelength, simulations are performed based on ray-tracing method using Zemax OpticStudio, a commercial software  \cite{zemax}. The simulated structure is shown in figure \ref{fig2}. As seen in figure \ref{fig2}, the simulated structure consists of a source, a sample, a beam splitter cube, a fixed and a movable mirror and a detector.    

An interferogram is generated by calculating the total intensity of rays reaching at the detector while the movable mirror is placed at different positions. The simulation is repeated for different positions of the moving mirror, and in each simulation the intensity of the electromagnetic wave arriving at the detector is calculated to form the interferogram. In each simulation, thirty million rays are launched. Fast Fourier transform (FFT) is applied to convert the interferogram data into the spectral intensity according to equation (\ref{eq.5}).

\begin{figure}
  \begin{subfigure}[b]{0.5\textwidth}

   \includegraphics[width=1\textwidth,trim=20px 0px 40px 40px,clip=true]{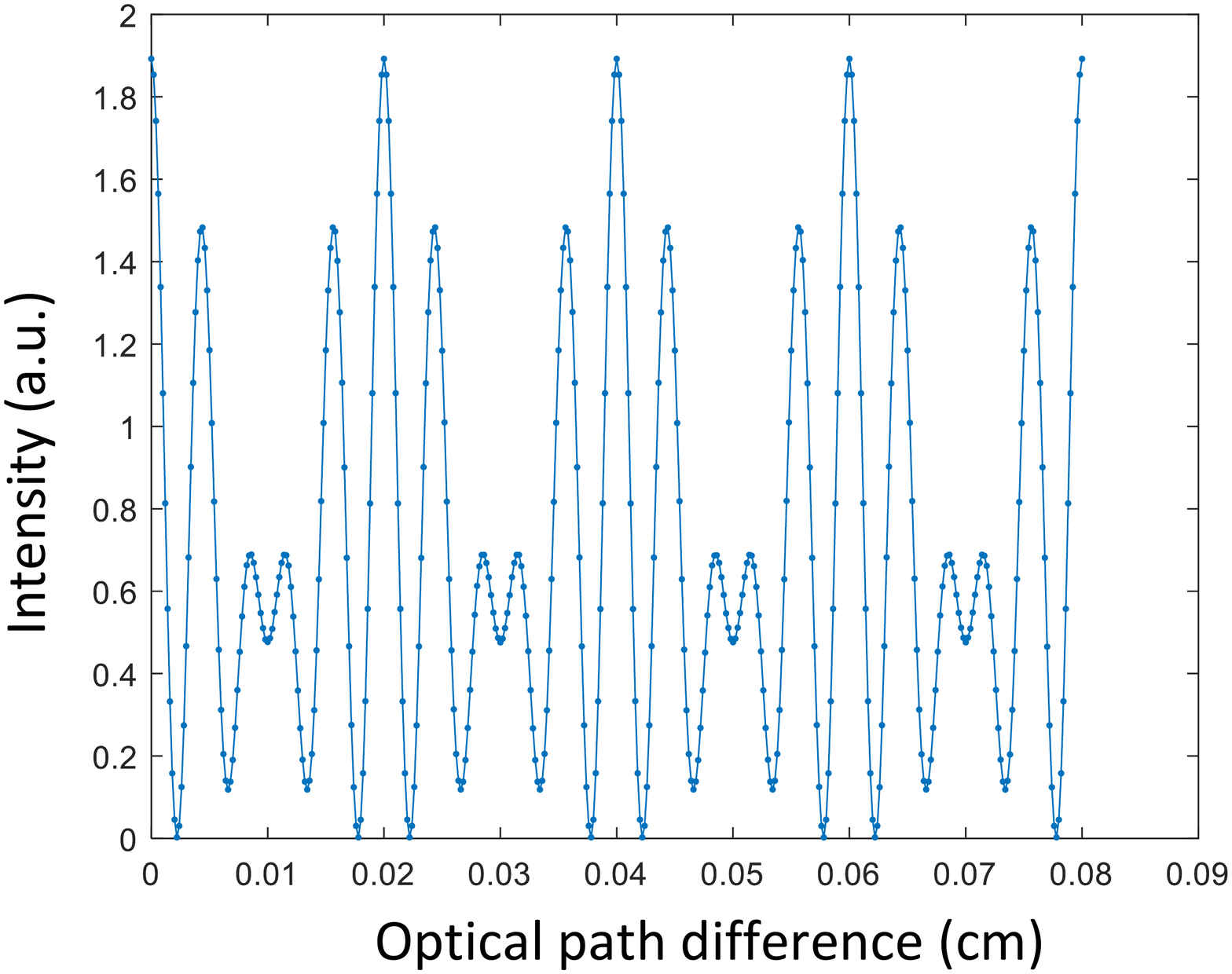}
    \caption{\label{fig3} }

    \end{subfigure}~
    \begin{subfigure}[b]{0.5\textwidth}

  \includegraphics[width=1\textwidth,trim=20px 0px 80px 20px,clip=true]{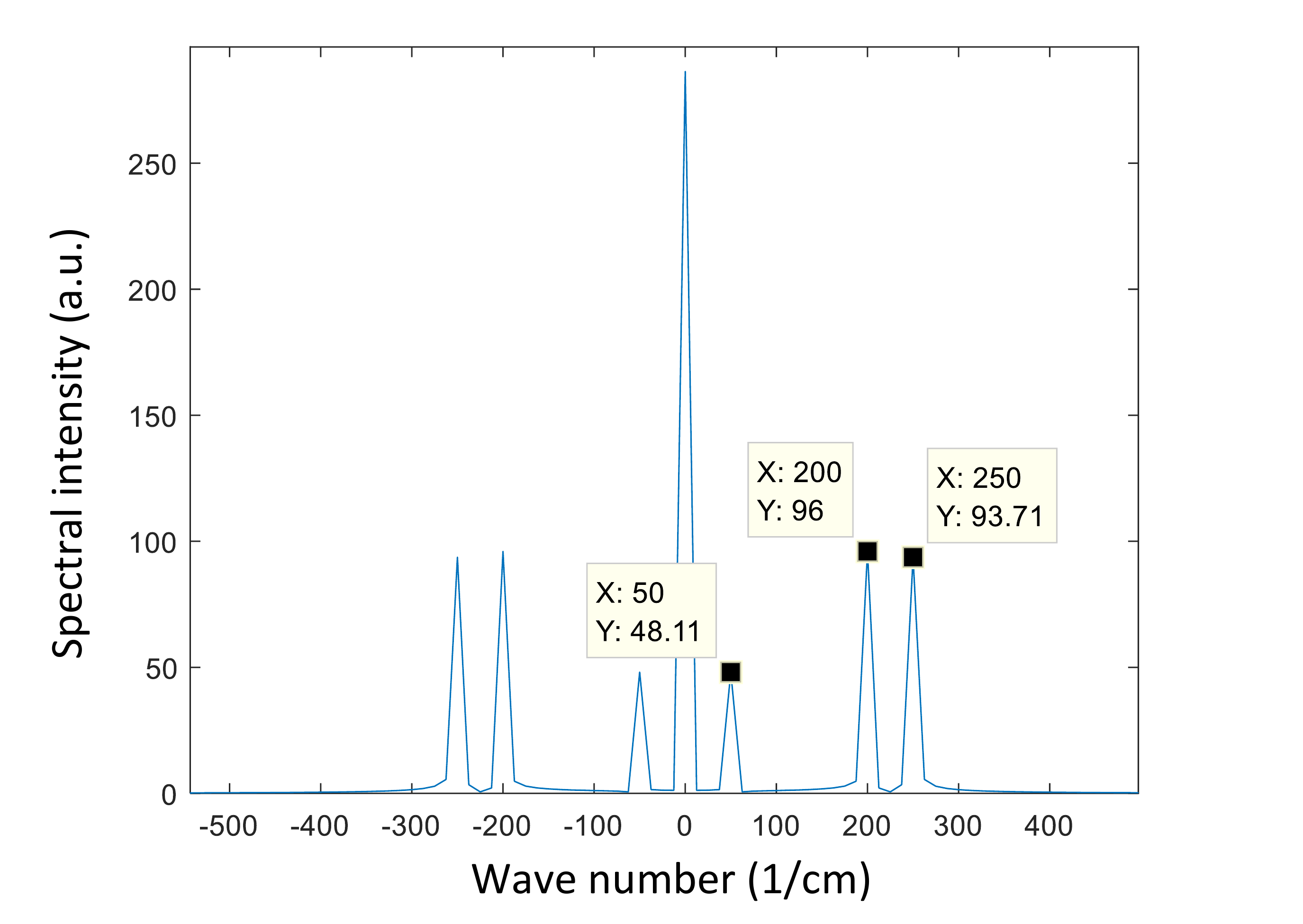}
    \caption{\label{fig4}}
    
    \end{subfigure}
    \caption{\label{fig34}(a) Simulated interferogram for an example with two-tone coherent source as described in the text. (b) FFT of the interferogram.}
    \end{figure}

In the first simulation, the source is assumed to have two distinct tones at wavenumbers of 200 cm$^{-1}$ and 250 cm$^{-1}$, and its coherent length is more than the maximum optical path difference used in the simulation. In this simulation, 401 discrete positions of the movable mirror with 8 $\mu$m displacement are considered. The simulated interferogram is shown in figure \ref{fig3}, whereas its FFT representing the spectral intensity can be seen in figure \ref{fig4}. 

\begin{figure}
\begin{center}
 \includegraphics[width=0.5\textwidth,trim=100px 10px 80px 80px,clip=true]{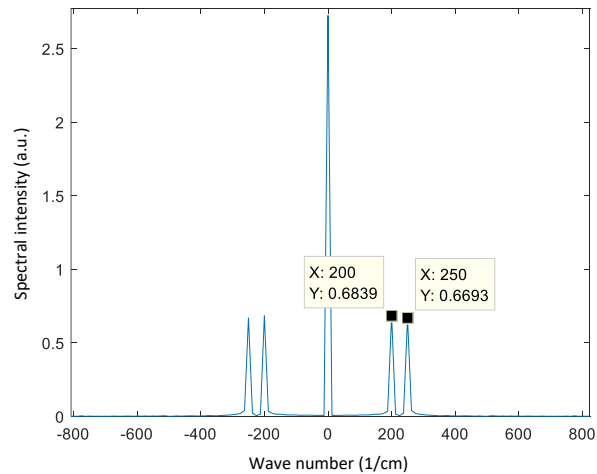}
\caption{\label{fig5}Simulated spectral intensity of an incoherent source with two distinct tones.}
\end{center}
\end{figure}

As it is evident in the calculated spectral intensity shown in figure \ref{fig34}, beside the source wavenumbers at 200 cm$^{-1}$ and 250 cm$^{-1}$,  a spurious signal at difference wavenumber of 50 cm$^{-1}$ appears in the spectrum. In the folowing reasoning, we will show that such difference wavenumber spurious signal is due the mixing effect of the two coherent tones in the detector. The total electric field of two coherent tones with equal amplitude, and its corresponding intensity at the detector can be modeled as 

 \begin{eqnarray}
 \label{eq.6}
   \mathbf{E}_R&=e^{-j k_1 z_1}+ e^{-j k_1 z_2}+e^{-j k_2 z_1}+e^{-j k_2 z_2}\\
   \label{eq.7}
   \mathbf{I}&=|\mathbf{E}_R|^2=\mathbf{E}_R\mathbf{E}_R^*\\
   \nonumber
   &=4+2\cos{k_1(z_2-z_1)}+2\cos{z_1(k_2-k_1)}+...\\
   \nonumber
   &...+2\cos{(k_1z_1-k_2z_2)}+2\cos{(k_1z_2-k_2z_1)}+...\\
   \nonumber
   &...+2\cos{z_2(k_2-k_1)}+2\cos{k_2(z_2-z_1)}
 \end{eqnarray}
 
\noindent where $k_1=2\pi\sigma_1$ and $k_2=2\pi\sigma_2$ are the propagation constant of the first and second tone, respectively, and $z_1$ and $z_2$ is the distance traveled by the wave in the fixed and movable mirror arms, respectively.
By inspecting equation (\ref{eq.7}), one would find out that the spurious signal at the difference wavenumber is due to the term given by $2\cos{z_2(k_2-k_1)}$. It should be noted that such cross terms will be averged out if the source has a coherent length of less than the optical path difference, or it becomes completely incoherent. The simulated spectral intensity with an incoherent source having the same two tones as in the previous simulation is shown in figure \ref{fig5}. It is evident that the spurious signal at the difference wavenumber is omitted in figure \ref{fig5}.

\begin{figure}
  
\begin{subfigure}[b]{0.5\textwidth}

   \includegraphics[width=1\textwidth,trim=20px 0 40px 0,clip=true]{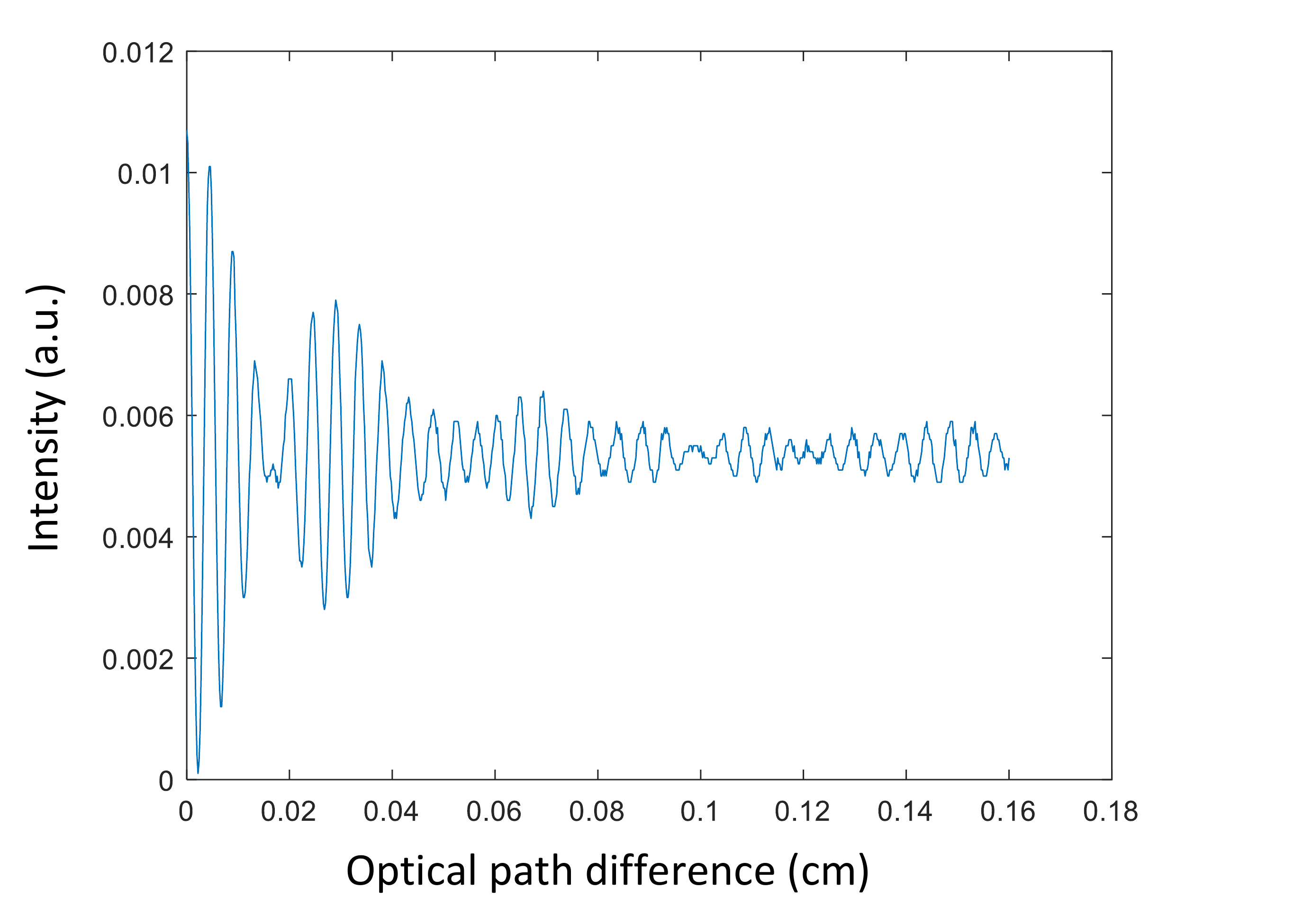}
    \caption{\label{fig7.1}}

    \end{subfigure}~
 \begin{subfigure}[b]{0.5\textwidth}

  \includegraphics[width=1\textwidth,trim=30px 30px 40px 0px,clip=true]{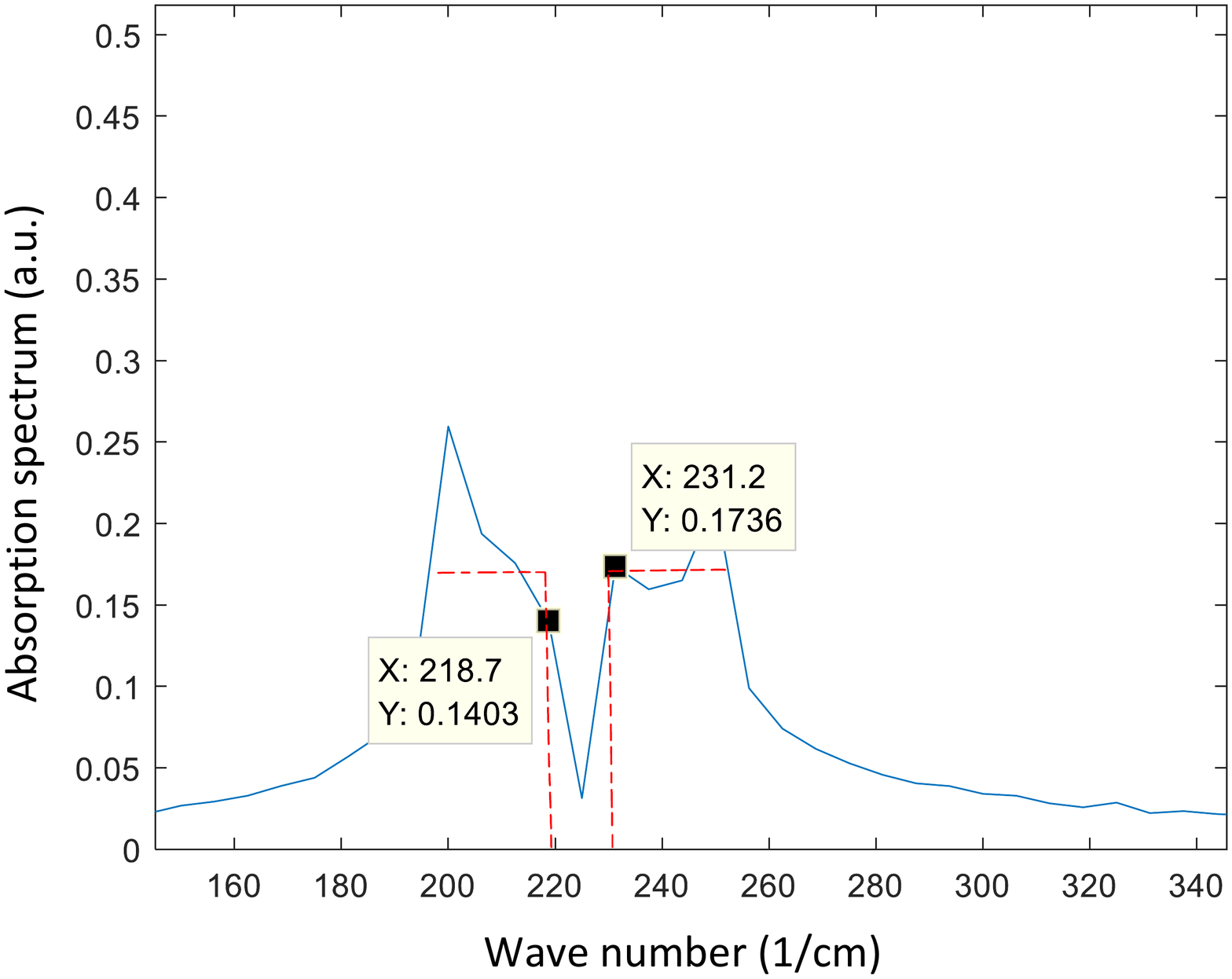}
    \caption{\label{fig7.2} }

    \end{subfigure}
    
 \caption{\label{fig7.3} (a) Simulated interferogram for an example with broadband incoherent source and an absorptive sample as described in the text. The maximum displacement of the moving mirror is 0.16 cm, (b) Recovered absorption spectrum of a sample (solid line) and its comparison with the absorption spectrum of the sample (dashed line) as described in the text.}
\end{figure}

In another simulation example, a broadband incoherent source with uniform spectral intensity covering the wavelength range of 40 $\mu$m to 50 $\mu$m is used. The goal is to simulate the FTS with a sample that has 100\% absorption within the wavelength range of 44 $\mu$m to 46 $\mu$m. The interferogram of this simulation is shown in figure \ref{fig7.1}. By taking FFT, the simulated spectral intensity is obtained in figure \ref{fig7.2}. 

\begin{figure}
  
\begin{subfigure}[b]{0.48\textwidth}

   \includegraphics[width=1\textwidth,trim=0px 0px 50px 0px,clip=true]{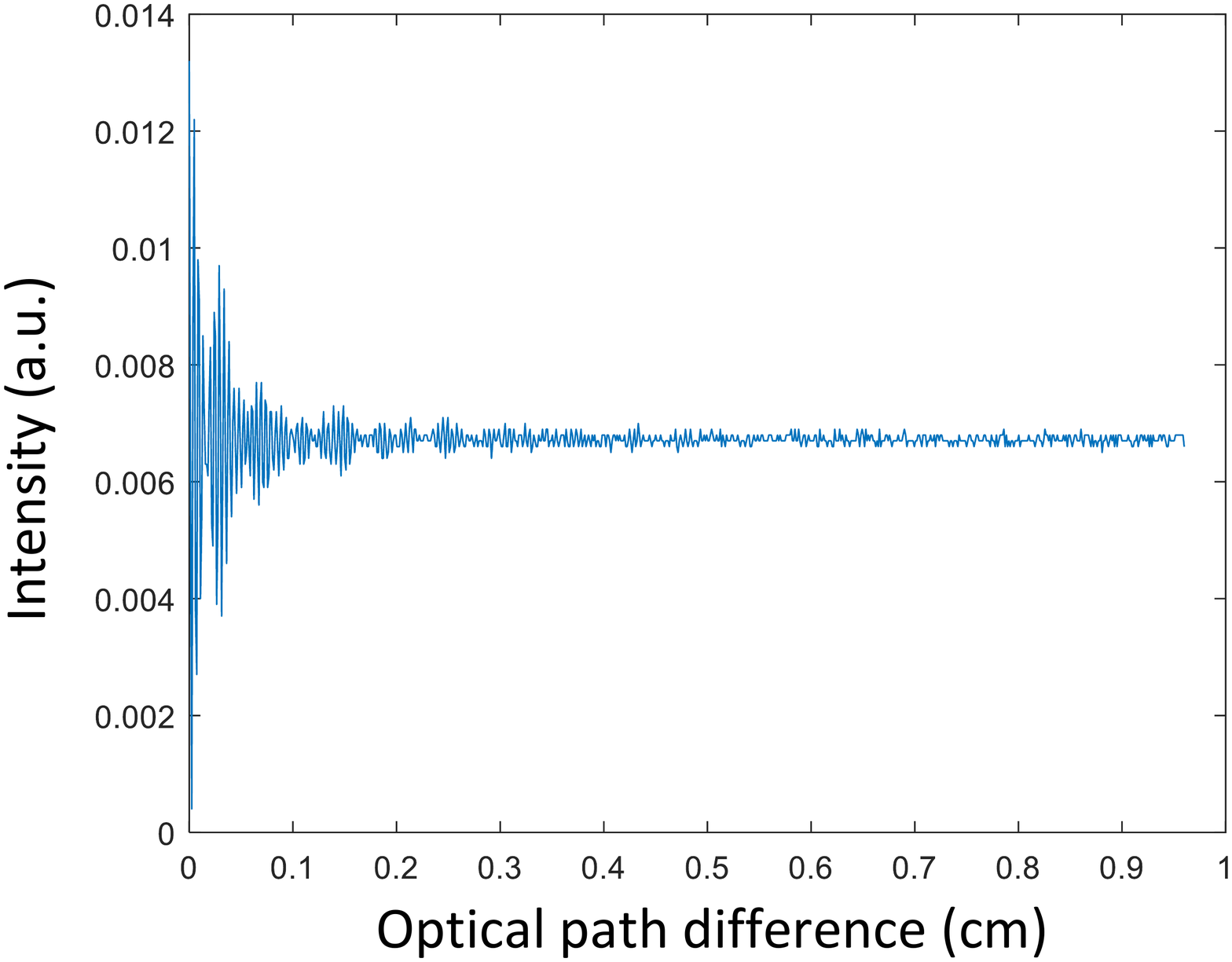}
    \caption{\label{fig6}}

    \end{subfigure}~
 \begin{subfigure}[b]{0.52\textwidth}

  \includegraphics[width=1\textwidth,trim=0px 30px 60px 0px,clip=true]{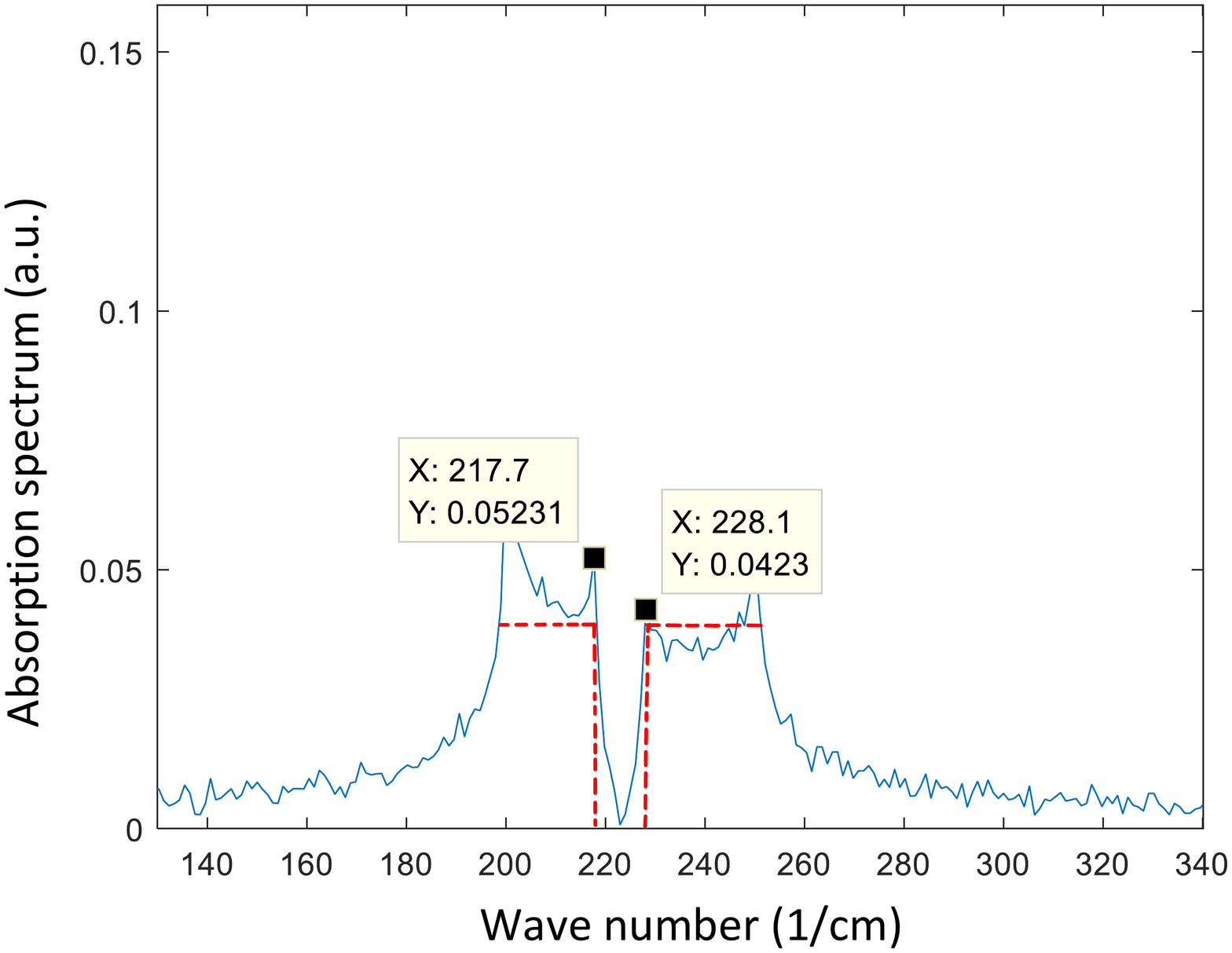}
    \caption{\label{fig7} }
\end{subfigure}
    
 \caption{\label{fig67} (a) Simulated interferogram for an example with broadband incoherent source and an absorptive sample as described in the text. The maximum displacement of the moving mirror is 0.95 cm, (b) Recovered absorption spectrum of a sample (solid line) and its comparison with the absorption spectrum of the sample (dashed line) as described in the text.}
\end{figure}

As it is evident in figure \ref{fig7.2}, there is a notch in the simulated spectrum, but the spectral  resolution is not good enough. By increasing the maximum displacement of the moving mirror, one can reach to the better spectral resolution \cite{bell2012introductory}. Figure \ref{fig6} shows the same interferogram, but for increased displacement of 0.95 cm as compared to 0.16 cm in figure \ref{fig7.1}. A more accurate spectrum is obtained in figure \ref{fig7}, such that there is a null between wavenumbers of 217.7 cm$^{-1}$ and 228.1 cm$^{-1}$ corresponding to the absorption spectrum edges of the sample, i.e. 44 $\mu$m to 46 $\mu$m. 

\begin{figure}
\begin{center}
\includegraphics[width=1\textwidth,trim=0px 40px 0 0,clip=true]{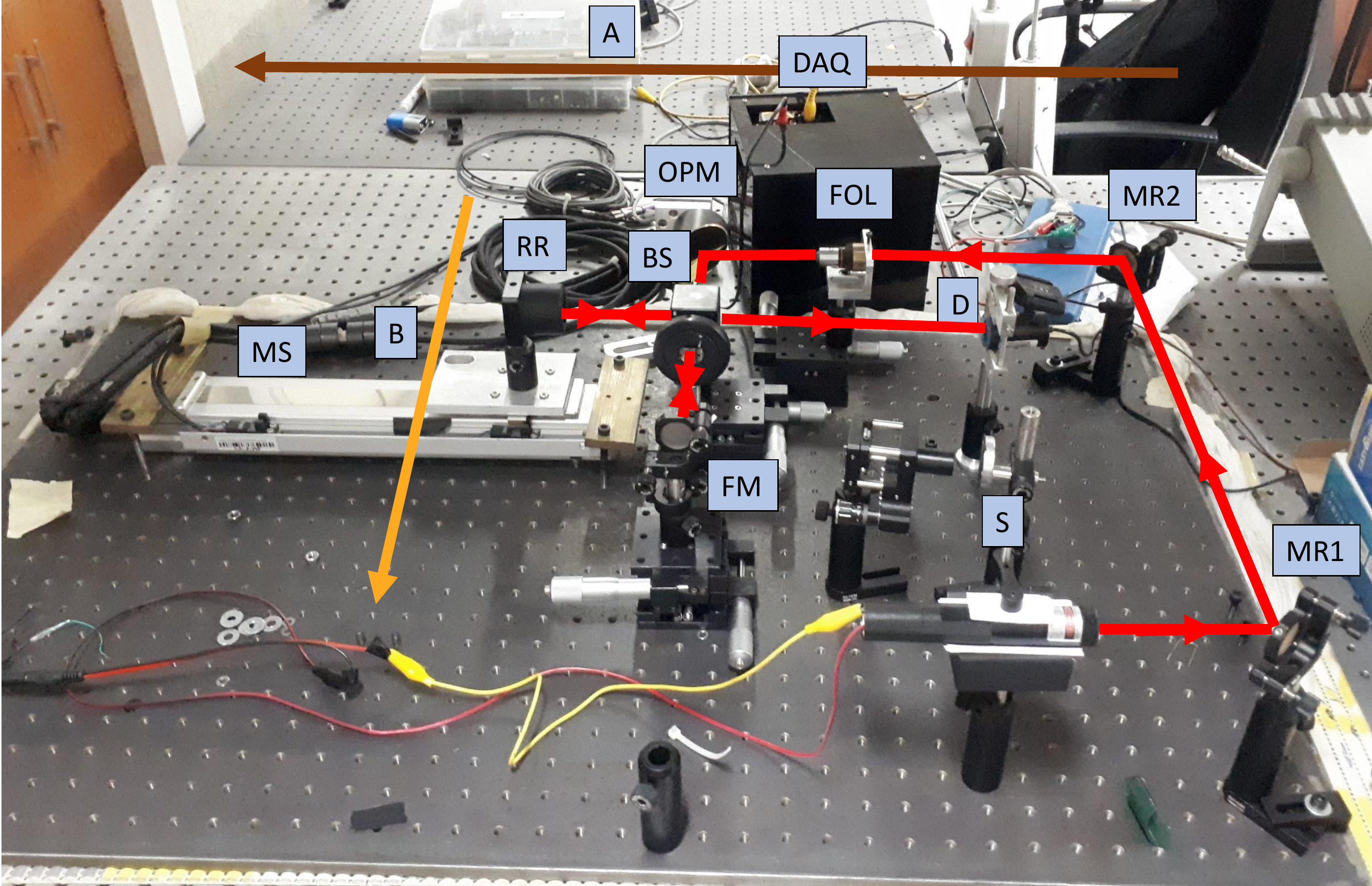}
\caption{\label{fig8}The implemented FTS setup based on Michelson interferometer (BS:Beam splitter, S:Source, D:Detector, RR:Retro-reflector, OPM:Off-axis parabolic mirror,
FM:Fixed mirror, MR:Aligning mirror, FOL:Focusing objective lens, MS:Motorized stage, DAQ:Data acquisition unit). The two arms of the Michelson interferometer are from BS to RR, and from BS to FM.}
\end{center}
\end{figure}

\begin{figure}
\begin{center}
\includegraphics[width=1\textwidth,trim=20px 70px 0px 70px,clip=true]{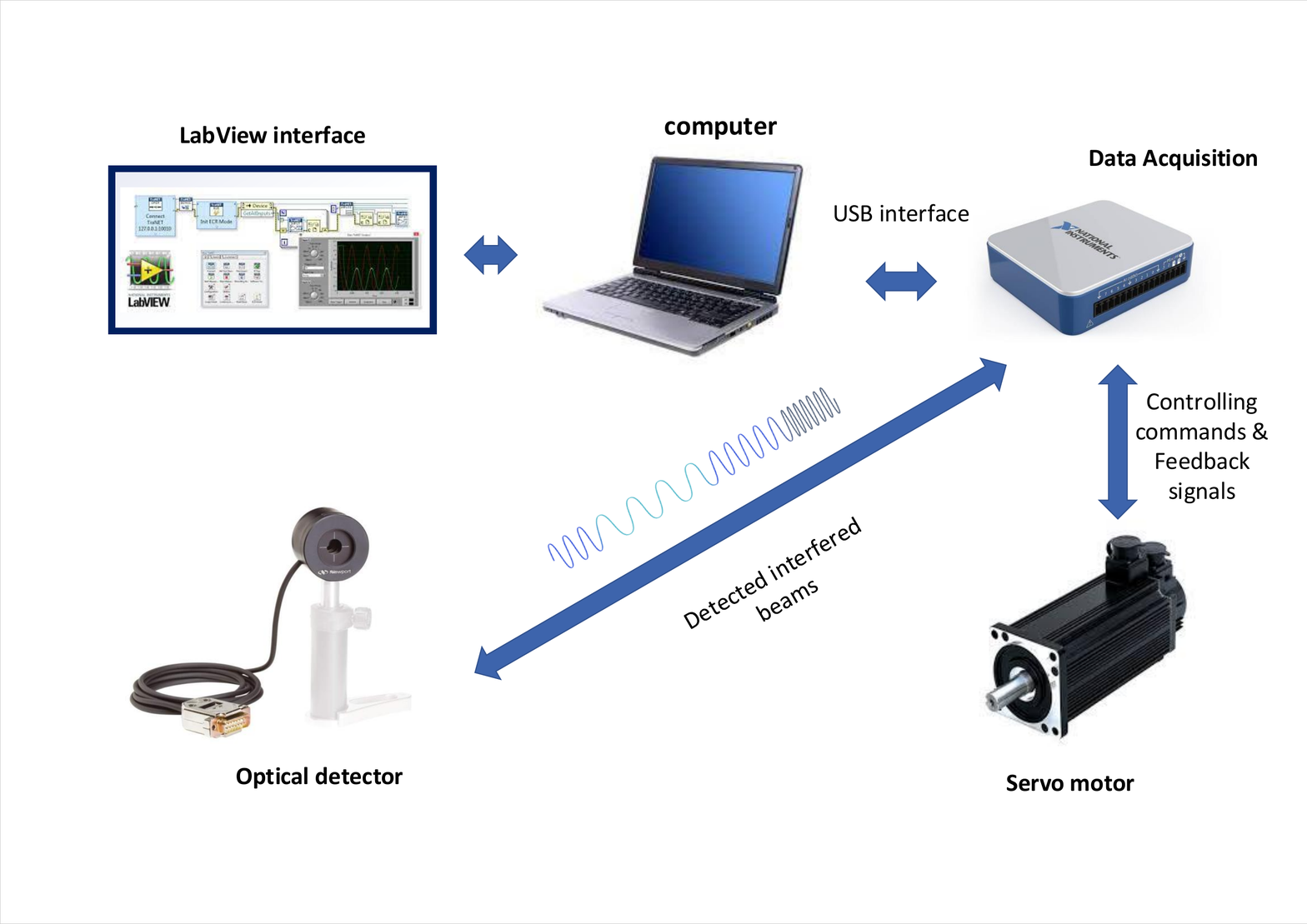}
\caption{\label{fig0} Schematic diagram of the connections between main parts of the FTS setup.}
\end{center}
\end{figure}

\begin{figure}
\begin{center}
\includegraphics[width=0.7\textwidth,trim=0 0 0 0,clip=true]{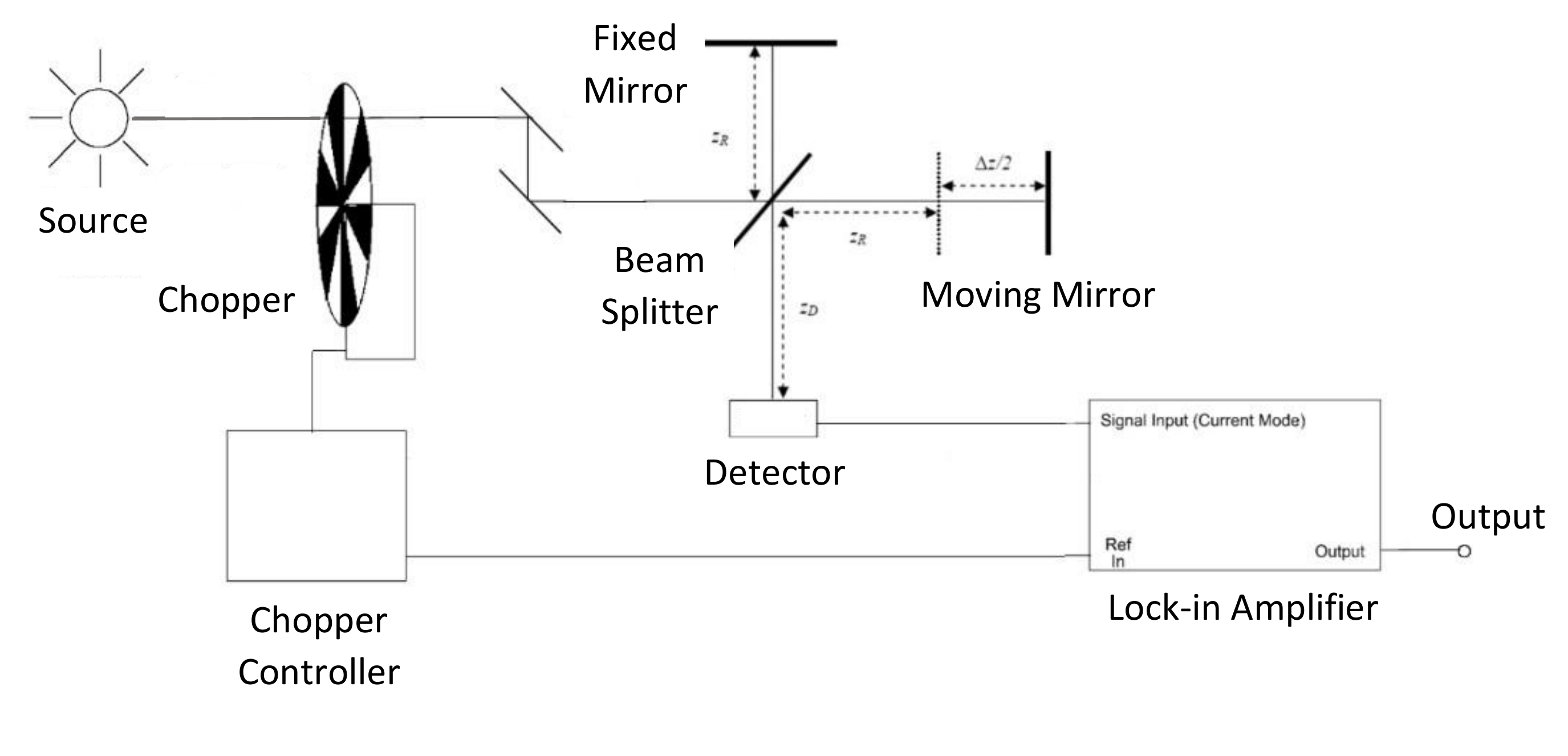}
\caption{\label{fig9} Michelson interferometer with chopper and lock-in amplifier.}
\end{center}
\end{figure}

\section{\label{implementation}FTS setup implementation}

In order to implement the FTS setup, we used optical components and equipment such as a source, a detector, mirrors, a retro-reflector, a beam splitter, a linear motorized stage, a multi-channel data acquisition (DAQ) unit, a chopper, a lock-in amplifier, and a computer to control the whole system. The implemented setup based on the Michelson interferometer is shown in figure \ref{fig8}.

A servo motor is used to move a linear stage on which a retro-reflector is mounted. The advantage of a retro-reflector over a flat mirror is that the reflected beam is always parallel to the incident beam, and it is ideally insensitive to the angular orientation. This makes the optical alignments much easier and less sensitive to the stage movements. As shown schematically in figure \ref{fig0}, a DAQ unit is used to send control data from the computer to the servo motor, e.g. to control its speed. Moreover, the position/velocity of the stage is extracted from the data received by the DAQ from an integrated encoder with the motor. A LabView program was developed to control the motor position and velocity by the computer through the DAQ in a closed loop fashion where the feedback is provided by a digital encoder. The DAQ unit also captures the output of the optical detector in order to create the interferogram, and digitizes the output for further processing by the computer. In the LabView program, the FFT of the captured interferogram is calculated and displayed. 

It is worth noting that during the data capturing from the detector output, the data points may not be recorded with the same intervals of motor displacement due to slight velocity fluctuation of the motor during normal speed or when the motor starts to accelerate/decelerate from/to the stop position. Therefore, it is necessary to re-sample the captured interferogram with equal sampling steps so that the FFT algorithm can be applied.

\subsection{\label{chopper}Chopper design and fabrication}

A lock-in amplifier is used to measure a very low signal, even lower than the unwanted interfering signals \cite{meade1982advances,scofield1994frequency}. The lock-in amplifier uses a technique so called phase-sensitive detection (PSD), which extracts the target signal with a specific phase and frequency from a contaminated signal with noise and interference. In the FTS setup, a lock-in amplifier can significantly reduce the noise and interference, e.g. from the motor circuitry, on the output of the detector. Figure \ref{fig9} illustrates an FTS setup using a lock-in amplifier and a chopper. The beam from the source passes through the chopper, and its amplitude is on-off modulated. The chopper is similar to a fan where its rotating blades block and unblock the beam with a chopping frequency or the so-called reference frequency. The chopped beam enters the Michelson interferometer, and the interfering beams reach to the detector. It is important to note the output signal of the detector is also on-off modulated with the reference frequency. The modulated signal from the output of the detector is fed to the input of the lock-in amplifier. 

\begin{figure}
  
\begin{subfigure}[b]{0.6\textwidth}

   \includegraphics[width=1\textwidth,trim=0 0 0 0,clip=true]{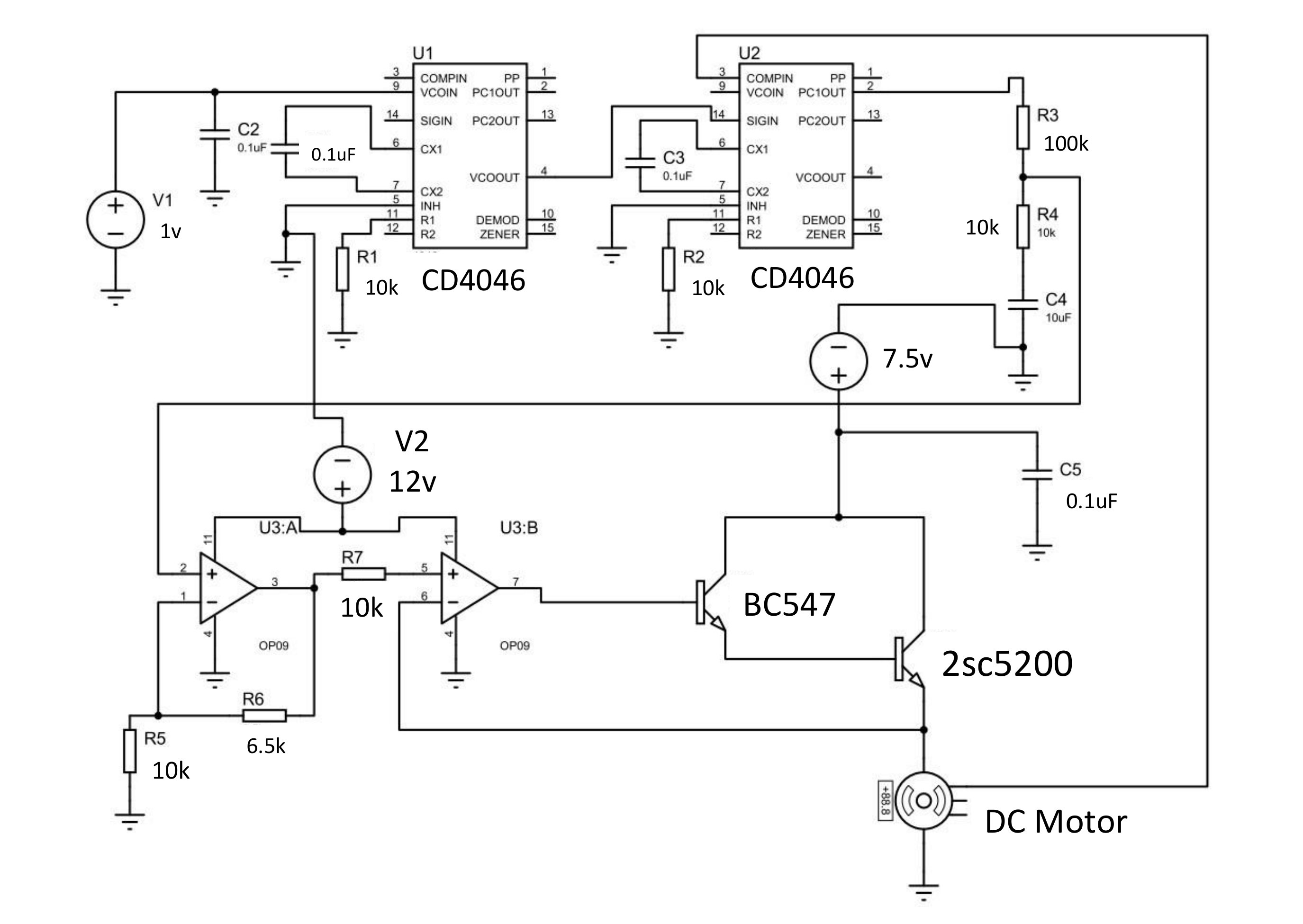}
    \caption{\label{fig10.2}}

    \end{subfigure}~
 \begin{subfigure}[b]{0.4\textwidth}

  \includegraphics[width=1\textwidth,trim=0px 120px 0px 0px,clip=true]{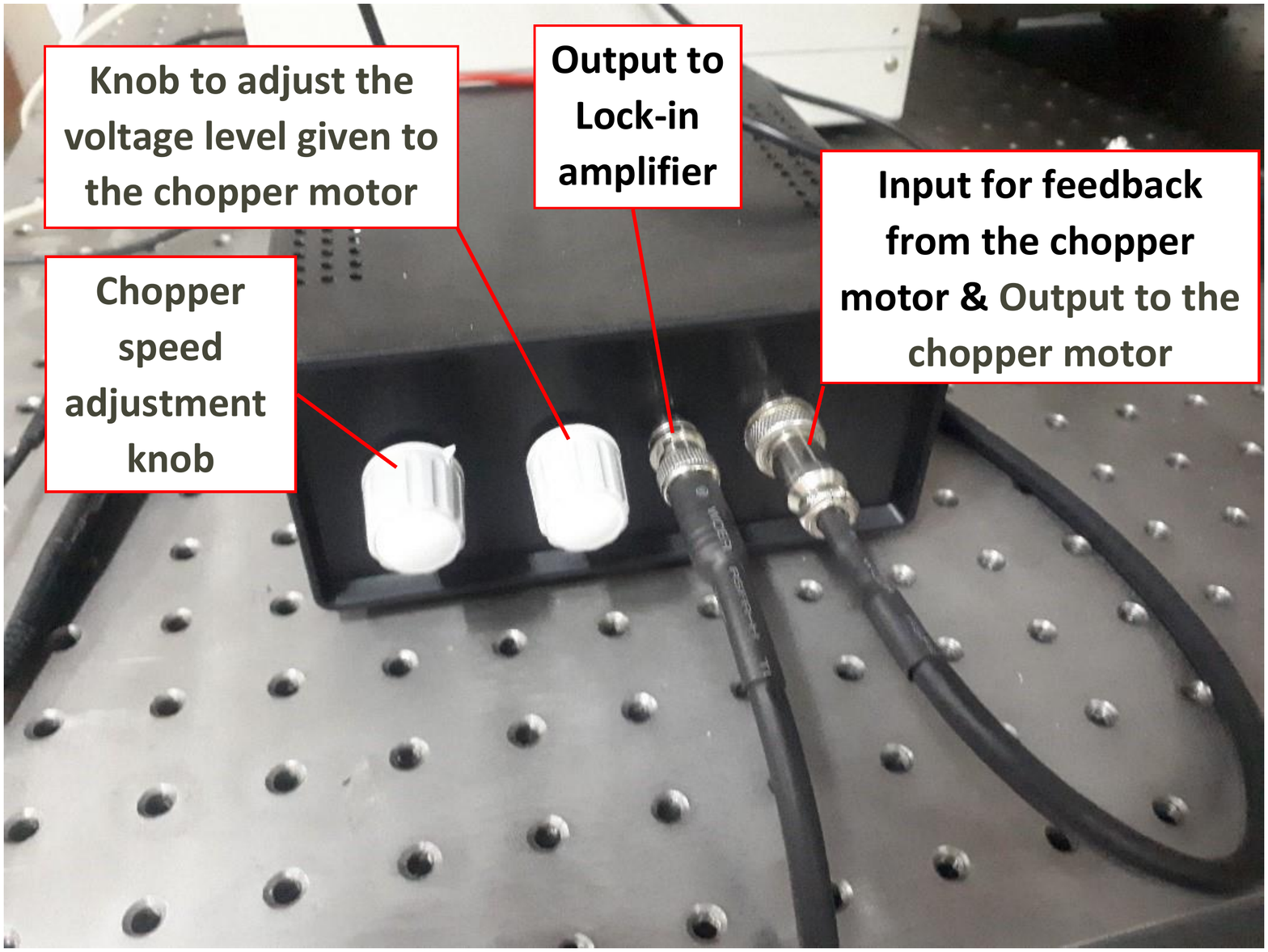}\vspace{1.5cm}
    \caption{\label{fig10.1} }

    \end{subfigure}
    
 \caption{\label{fig10} (a) Schematic of the chopper controller circuit, (b) fabricated chopper.}

 \end{figure}

The chopper has a controller through which the user can set the reference frequency. The reference frequency is proportional to the speed of the chopper. The speed of the chopper motor is controlled in a closed loop fashion from a feedback signal provided by an optical encoder. The chopper controller also generates a signal with the exact frequency of the chopped beam that is fed to the reference input of the lock-in amplifier. Figure \ref{fig10} shows the schematic of the chopper controller circuit and a photo of its implementation. The main components are two digital VCOs with part number CD4046 and a DC motor as shown in figure \ref{fig10.2}. The first VCO designated by U1 in the schematic, sets the reference frequency through a voltage value, and the second VCO designated by U2 is used in a closed loop to control the chopper frequency, and keep it constant during the normal operation. A low pass filter is used to provide the driving DC voltage of the motor from the output of the second VCO. Upon locking of the control loop, the feedback signal from the optical encoder provides the reference frequency for the lock-in amplifier.

The lock-in amplifier has a frequency locking loop that locks into the reference signal frequency and creates an internal reference signal with a frequency exactly equal to the input reference. The lock-in amplifier first amplifies the input signal fed from the detector, and then mixes it by in-phase and quadrature (I/Q) components of the sinusoidal signal generated from the reference signal. The I/Q demodulated signals are then filtered by low-pass filters to extract the original target signal from the unwanted noise or any other interference. It should be emphasized that since the noise or any other interference is not modulated with the chopping frequency, it is suppressed by the low pass filtering after demodulation in the lock-in amplifier.

\subsection{\label{detector}Circuitry of optical detector}
\begin{figure}
\begin{center}
\includegraphics[width=0.8\textwidth,trim=0 0 0 0px,clip=true]{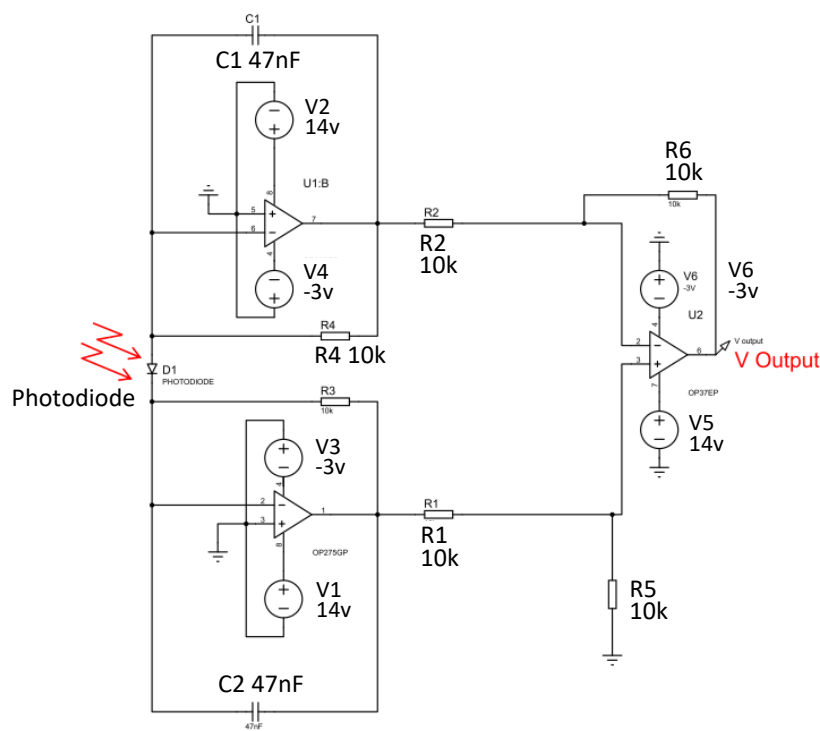}
\caption{\label{fig11} Schematic circuit of the filtering buffers and the amplifier for the photodiode.}

\end{center}
\end{figure}

A photodiode is used to convert the optical intensity into the electrical signal. However, since there are various sources that can produce noise and interference, such as servomotor and unwanted background light, the output current of the photodiode needs filtering and amplification. Figure \ref{fig11} shows the schematic circuit used to filter and amplify the output of the detector. Two buffers are used as low pass filters to reduce the detector noise where their cut off frequencies are adjusted to pass the main signal and suppress the excess noise. The bias voltages of op-amps are set such that to avoid output saturation at the maximum optical intensity.

\subsection{\label{alignment}Optical alignment}
The optical alignment plays a crucial role in the proper operation of the FTS. The goal in optical alignments is mostly to guide the beam on a straight path as parallel to the surface of the optical table as possible.
The azimuth and elevation of the beam can be adjusted by two mirrors mounted on kinematic mounts in an iterative procedure \cite{bell2012introductory}. In figure \ref{fig8}, MR1 and MR2 are used to align the beam along the path shown by arrow 'A'. To do this, two pinholes are first placed in path 'A' that are one meter apart along one row of tapped holes in the optical table. Then, by adjusting the azimuth and elevation of the beam through the knobs on the kinematic mounts, an attempt is made to pass the aligning beam through the center of both pinholes. In the alignment procedure, when MR1 is adjusted, the attempt should be made to pass the beam through the center of the pinhole closer to MR2. When MR2 is adjusted, the attempt should be made to pass the beam through the center of the pinhole farther from MR2, and this procedure should be repeated back and forth a few times so that the desired alignment is achieved. Since adjusting exactly the same height for two pinholes may not be easy, it is suggested that one pinhole with fixed height is switched between two positions during the above procedure.

In the next step, we follow the same procedure by adjusting the orientation of the focusing lens shown as FOL in figure \ref{fig8}, and the off-axis parabolic mirror for alignment in the direction shown by arrow 'B' so that the beam hits as perpendicular to the beam splitter cube surface.
With a good alignment, one would get an interference pattern that has wide fringe lines at the detector. Ideally, the interference pattern of two completely colinear plane waves is uniform without any fringes. Therefore, by inspecting the interference fringes, one should try to make them as wide as possible through adjusting the kinematic mount of fixed mirror and off-axis mirror, and the manual stages as shown in figure \ref{fig8}. Such attempt leads the two interfering beams to become as colinear as possible.

Our setup was tested using a laser source with its maximum power around the wavelength of 808 nm, i.e. equivalent to wavenumber of 12376 cm$^{-1}$. In order to access an inexpensive laser source, we used a commercial green laser pointer, and removed the conversion wavelength crystals from its output head. Also we used a low cost webcam, and removed its front lens to reveal its array detector for recording the fringe lines at the output path of the Michelson interferometer. 

\begin{figure}
 
\begin{subfigure}[b]{0.4\textwidth}

   \includegraphics[width=1\textwidth,angle=0,trim=0 0 0 0px,clip=true]{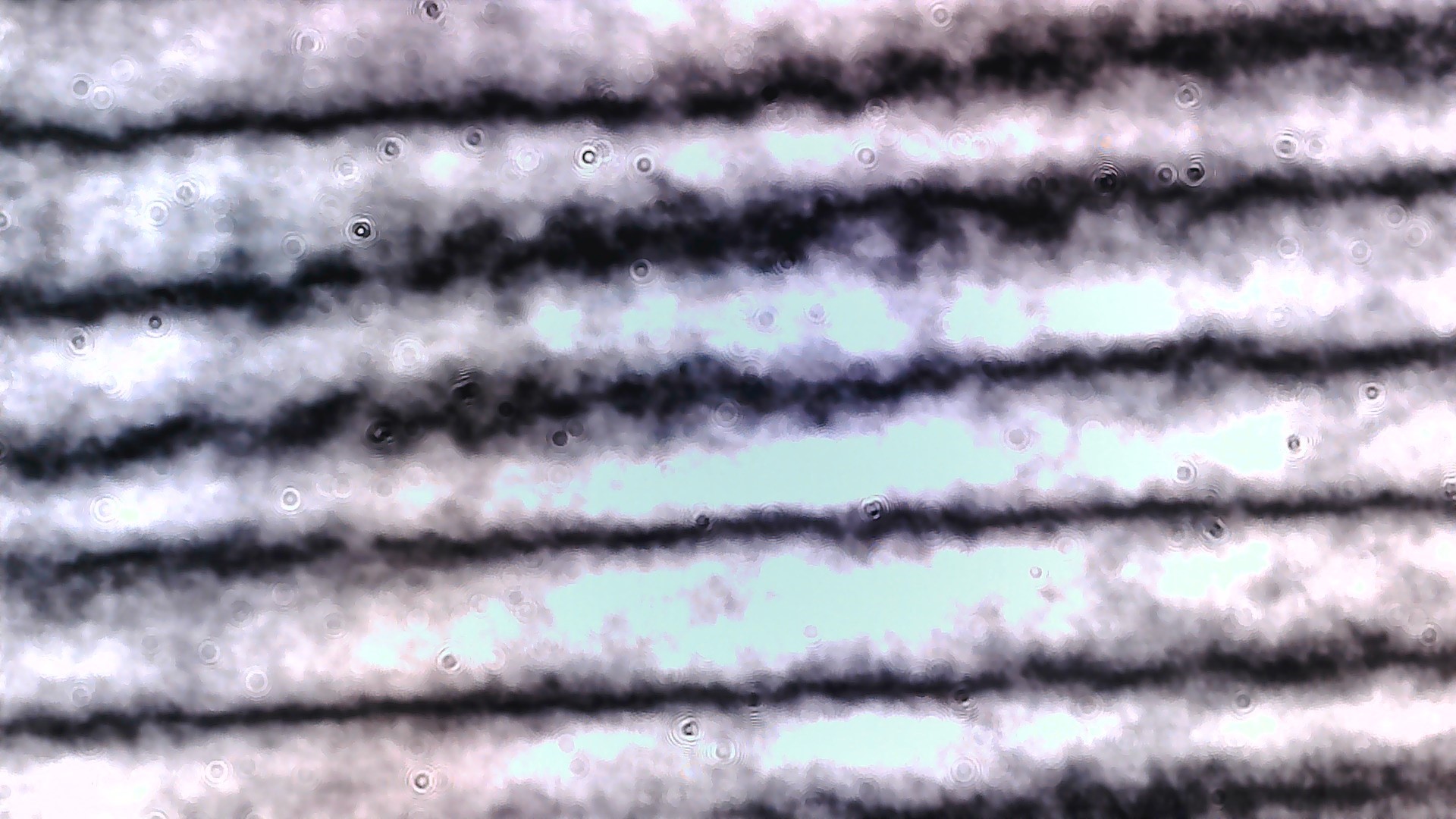}\vspace{+1.9cm}
    \caption{\label{fig12a}}

    \end{subfigure}~  
 \begin{subfigure}[b]{0.6\textwidth}

  \includegraphics[width=1\textwidth,trim=0px 0px 0px 0px,clip=true]{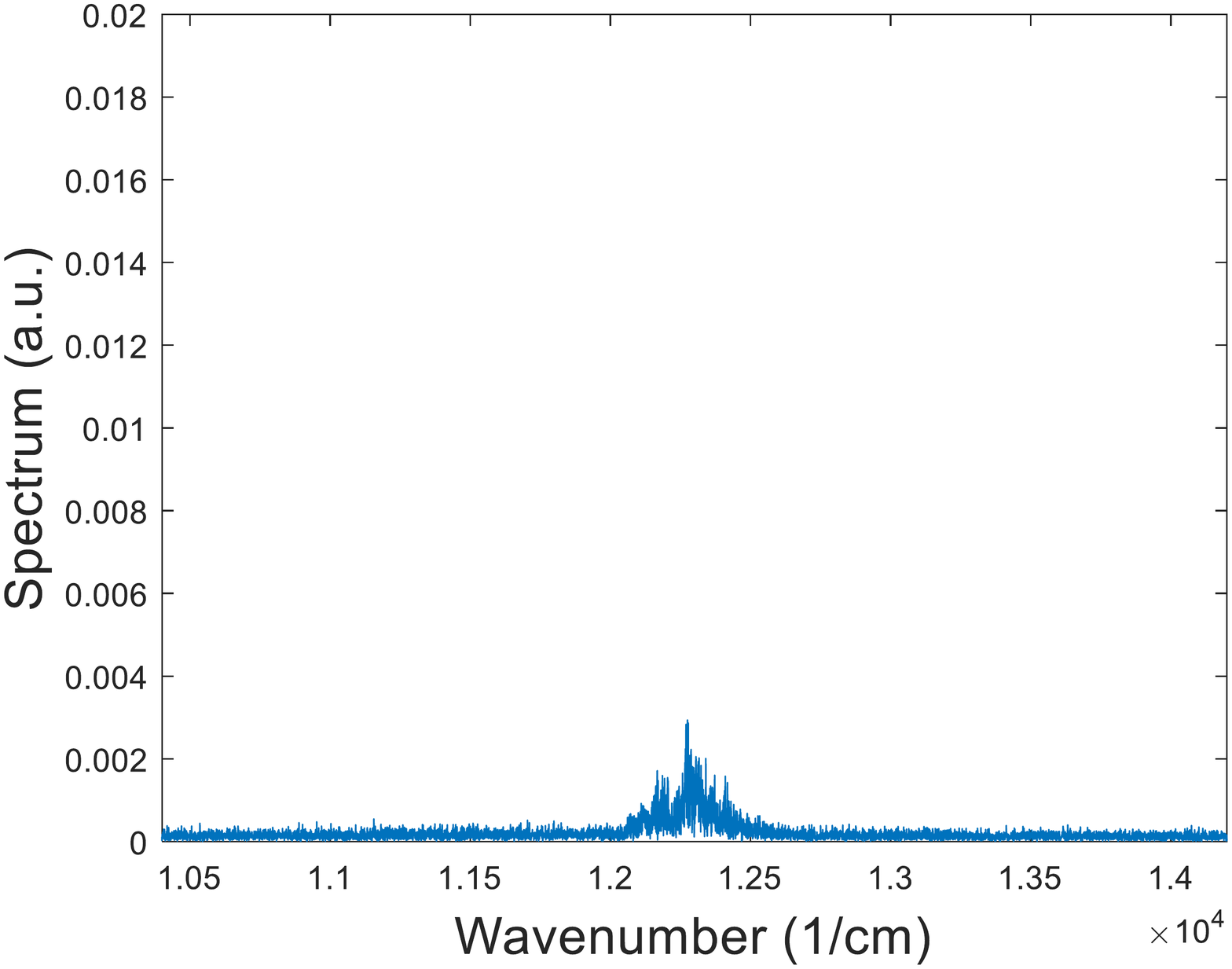}
    \caption{\label{fig12b}}

    \end{subfigure}
    
 \caption{\label{fig12} (a) Interference fringe line pattern in a poor alignment (b) the measured spectral intensity.}

 \end{figure}

\begin{figure}
 
\begin{subfigure}[b]{0.45\textwidth}

   \includegraphics[width=1\textwidth,angle=0,trim=0 0 0 0px,clip=true]{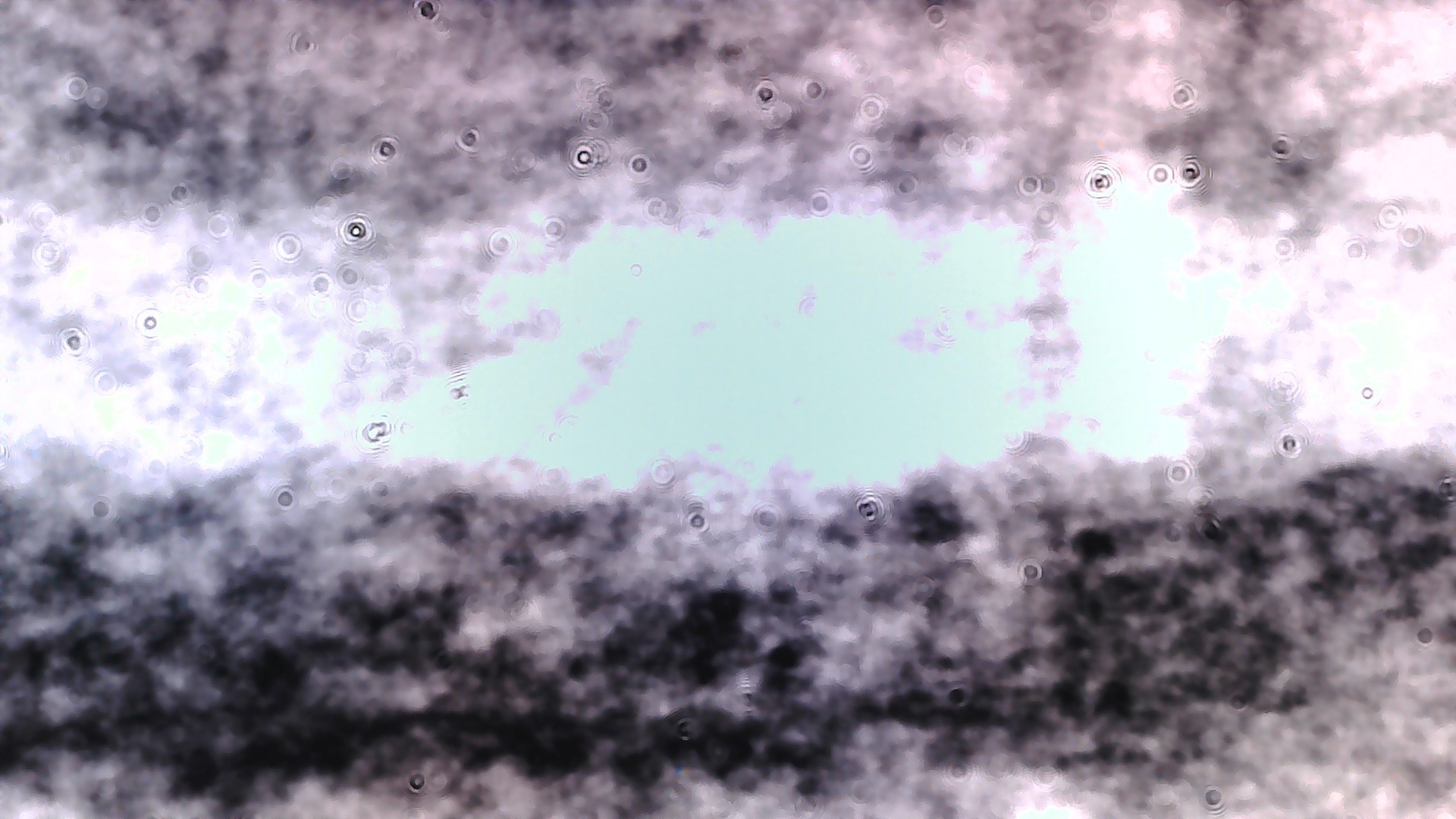}\vspace{+1.5cm}
    \caption{\label{fig13a}}

    \end{subfigure}~  
 \begin{subfigure}[b]{0.55\textwidth}

  \includegraphics[width=1\textwidth,trim=0px 0px 0px 0px,clip=true]{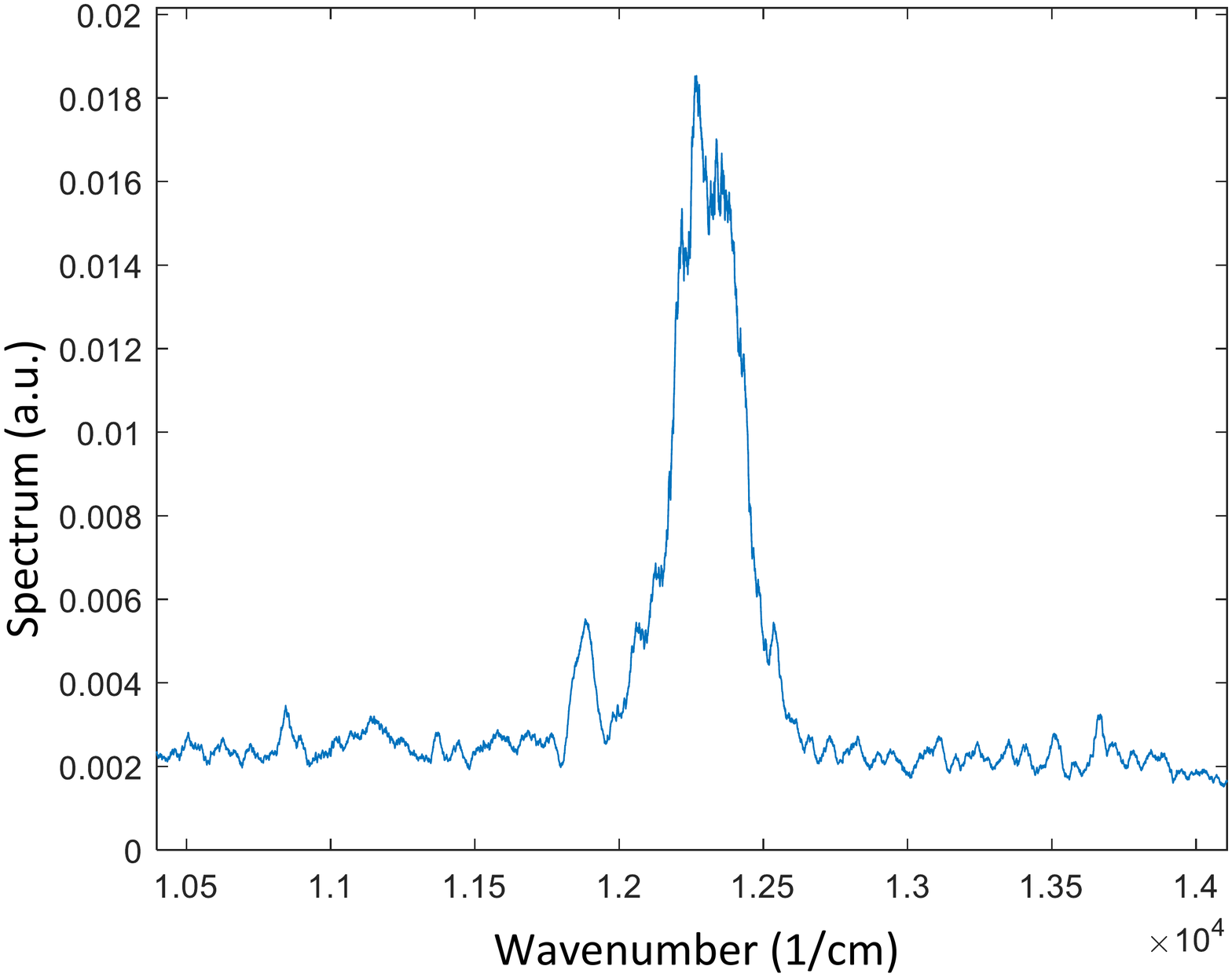}
    \caption{\label{fig13b}}

    \end{subfigure}
    
 \caption{\label{fig13} (a) Interference fringe line pattern in a good alignment (b) the measured spectral intensity.}

 \end{figure}

The fringe lines for two different optical alignments are shown in figures \ref{fig12} and \ref{fig13}. In order to record the interferogram, the array detector is replaced by the photodiode described above. Comparing measurements results in figures \ref{fig12b} and \ref{fig13b}, it is evident that how the measured spectral intensity is sensitive to optical alignment in an FTS based on Michelson interferometer. It is clear that the signal-to-noise ratio is enhanced considerably for a good alignment. In figure \ref{fig13b}, the peak of the spectrum coincides with a wavenumber that is equivalent to 808 nm as expected.

\section{Conclusion}
Various aspects of a Fourier transform spectroscopy setup based on Michelson interferometer were studied. On practical side, we realized that the noise and interference can be considerably suppressed by using proper filtering and utilizing the chopper and lock-in amplifier. Moreover, proper alignment is of paramount importance in the accuracy and signal-to-noise ratio of the measured spectral intensity. Aside from many applications of FTS in  chemistry, food and pharmaceutical industry, and environmental studies to name a few, implementing its apparatus with inexpensive laboratory items is a powerful educational tool for physics and electrical engineering students.  
\section*{References}
\bibliographystyle{iopart-num}
\bibliography{FTS.bib}
\end{document}